\documentclass[doublecol]{epl2} 

\usepackage{verbatim}

\title{Stability of three-fermion clusters with finite range of attraction}
\shorttitle{Stability of three-fermion clusters} 

\author{Pavel E. Kornilovitch}
\shortauthor{P. E. Kornilovitch}

\institute{                    
  Hewlett-Packard Company -- Corvallis, OR, USA
}

\abstract{
Three quantum particles with on-site repulsion and nearest-neighbour attraction on a one-dimensional lattice are considered. The three-body Schr\"{o}dinger equation is reduced to a set of single-variable integral equations. Energies of three-particle bound complexes (trions) are found from self-consistency of the approximating matrix equation. In the case of spin-$\frac{1}{2}$ fermions, the ground state trion energy, the excited state energies, the trion spectra and stability regions are obtained for total spins $S = 1/2$ and $S = 3/2$. In the $S = 1/2$ sector, a narrow but finite parameter region is identified where the ground state consists of a stable fermion pair and an unbound fermion. Also presented is the reference case of spin-0 bosons.   
}

\begin{document}

\maketitle

\section{\label{3UV1d:sec:one}
Introduction}

It has been known for a long time that three spin-$\frac{1}{2}$ fermions interacting via point-like pair-wise attraction do not form three-particle bound clusters (trions) on one-dimensional~\cite{Takahashi1970} and two-dimensional lattices~\cite{Mattis1984,Rudin1985}. The situation is drastically different if the attractive potential has a finite radius. In this case, the exclusion principle does not prevent configurations where all three particles interact at the same time. As a result, trions are expected to form at least in the strong-coupling limit. If the two-body potential has a repulsive core, the attraction has to exceed a threshold value to create two-particle and three-particle bound states. The important question then is the order in which pairs and trions are formed and whether a parameter region exists where the trion is unstable against decay into a stable pair and a free fermion. In this paper, this question is answered for a one-dimensional lattice model with on-site (Hubbard) repulsion $U$ and nearest-neighbour attraction $V$.   

Motivation for this type of models comes primarily from theories of superconductivity. Already Mattis and Rudin noted the correspondence between Cooper pairs in 3D and low density fermion pairs in 2D~\cite{Mattis1984}. After the discovery of high-$T_c$ superconductors, Alexandrov and others~\cite{Micnas1990,Alexandrov1994,Alexandrov2010} promoted the idea that the ground state of doped cuprates is a Bose-liquid of fermionic pairs, glued together by a strong electron-phonon or mixed interaction. For these ideas to be valid, fermionic trions should not form before the pairs. Three-body lattice states also appear in exciton models~\cite{Shibata1990,Bulatov1994}, scattering of bound pairs off defects~\cite{Bulatov2005,Zhang2013} and in the description of ion traps~\cite{Petrov2012}. 

Early work on three-body lattice problems was reviewed by Mattis~\cite{Mattis1986}. Most of those and subsequent~\cite{Fabrizio1991,Valiente2010,Pohlmann2013} papers were confined to point-like interactions. There is only a handful of studies in which finite-radius interactions were considered. Rudin considered three spin-0 bosons in a $UV$ model identical to ours~\cite{Rudin1986}. He analyzed the trion mass in the strong-coupling limit but did not present the overall phase diagram. Shibata et al studied an exciton model with onsite repulsion between two holes and intersite attraction between the holes and an electron~\cite{Shibata1990}. Berciu presented a numerical method for spinless fermions with nearest neighbour and next nearest neighbour interaction~\cite{Berciu2011}.

\section{\label{3UV1d:sec:two} 
Model}

The model is described by the following one-dimensional $UV$ Hamiltonian 
\begin{equation}
H = - t \sum_{i, \delta = \pm 1} a^{\dagger}_i a_{i+\delta} + 
\frac{U}{2} \sum_i n_i ( n_i - 1 ) - V \sum_{i} n_i n_{i+1}  \: . 
\label{3UV1d:eq:one}
\end{equation}
Here $n_i$ is the number density operator for site $i$, and $\delta$ denotes nearest neighbours. Mathematically, the model is well defined for any values of $U$ and $V$ but physically the most interesting region is $U \gg t$ and $V \approx t$. The present study is limited to the $(U,V) > 0$ domain. The nearest-neighbour attraction of eq.~(\ref{3UV1d:eq:one}) might be regarded as a particular case of a wider class of small-depth finite-range attractive potentials~\cite{Kornilovitch1995,Valiente2010b}.    

The two-body problem in model~(\ref{3UV1d:eq:one}) can be solved exactly, see, e.g., the appendix in \cite{Kornilovitch2004}. Two spin-$\frac{1}{2}$ fermions form a singlet pair at $V > 2 U t \cos{(K/2)}/[U + 4t\cos{(K/2)}]$ and a triplet pair at $V > 2t \cos{(K/2)}$, where $K$ is the total lattice momentum of the pair. Only the first state is allowed for spin-0 bosons. The two-body $UV$ model has also been solved on the 2D square lattice for nearest-neighbour~\cite{Lin1991,Petukhov1992,Kornilovitch2004} and long-range attractions~\cite{Kornilovitch1995}.

\begin{figure}[t]
\begin{center}
\includegraphics[width=0.48\textwidth]{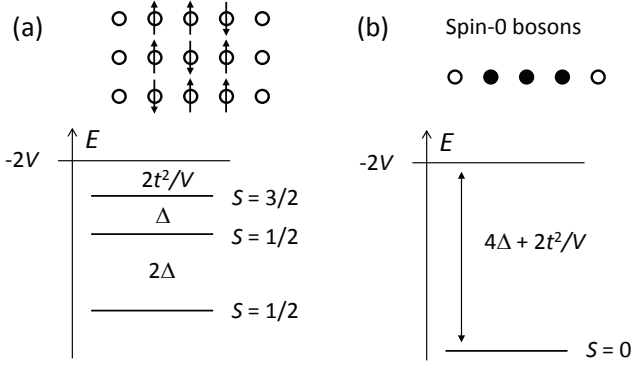}
\end{center}
\caption{In the $(U,V) \gg t$ limit, three particles occupy three adjacent lattice sites. (a) Schematic energy diagram of spin-$\frac{1}{2}$ fermions. When mixed by second-order hopping events, the three basic configurations (indicated by arrows) split into two $S = 1/2$ trions and one $S = 3/2$ trion. Each states is $(2S+1)$ degenerate. (b) In the case of spin-0 bosons, there is only one $S = 0$ trion with a fully symmetrical wave function.  $\Delta = t^2/U + t^2/(2V+U)$.} 
\label{3UV1d:fig:zero}
\end{figure}

It is instructive to begin analysis of three-particle states with the strong-coupling limit $(U,V) \gg t$. The particles occupy three adjacent lattice sites, with the strong $U$ term preventing double occupancy. Since only two attractive bonds are at work, in the leading order the trion energy is $E^{(0)}_3 = - 2 V$. In the case of spin-$\frac{1}{2}$ fermions, three basic configurations are shown in fig.~\ref{3UV1d:fig:zero}(a). When mixed by hopping events, the states split into two $S = 1/2$ trions and one $S = 3/2$ trion, as indicated on the diagram. Within the second-order perturbation theory in $t$, the trion energies are
\begin{equation}
E^{(2)}_3 = - 2V - \frac{2 t^2}{V} - m \left( \frac{t^2}{U} + \frac{t^2}{2V + U} \right) , 
\label{3UV1d:eq:two}
\end{equation}
where $m = 3, 1$ for the $S = 1/2$ states and $m = 0$ for the $S = 3/2$ state. In addition, the $S = 1/2$ trions are 2-fold degenerate whereas the $S = 3/2$ trion is 4-fold degenerate.    

The $S = 3/2$ trion state corresponds to the three-row Young diagram and is described by a fully antisymmetrical coordinate wave function. Both $S = 1/2$ states correspond to the two-row Young diagram and have wave functions of mixed symmetry. However, the coordinate wave functions of $S = 1/2$ trions can always be chosen symmetrical or antisymmetrical with respect to permutation of one pair of coordinates, for example particles 1 and 2. This property will form the basis of a numerical method described below.   

In three-body problems, spin-0 bosons provide a useful reference since their bound complex has the lowest energy among all possible trion state. In this paper, spin-0 zero bosons will be considered alongside spin-$\frac{1}{2}$ fermions. The strong coupling energy diagram is shown in fig.~\ref{3UV1d:fig:zero}(b). The energy of the sole state is given by eq.~(\ref{3UV1d:eq:two}) with $m = 4$. Hereafter, this state will be referred to as the $S = 0$ trion. It has a fully symmetrical coordinate wave function.   

If $V$ is lowered toward the $V \approx t$ region, the energy pattern described above gets distorted by the growing kinetic energy, and eventually the states begin to disappear into the pair plus free particle continuum.

\section{\label{3UV1d:sec:three}
Method}

The method used in this work is an extension of ref.~\cite{Mattis1984} to finite-range potentials. The three-particle Schr\"odinger equation $H \psi = E \psi$ in momentum space reads   
\begin{eqnarray}
& & [ E - \varepsilon(q_1) - \varepsilon(q_2) - \varepsilon(q_3)] \psi(q_1,q_2,q_3) = 
\nonumber \\
& & \frac{1}{N} \sum_k {\cal V}(k) \left[ \psi(q_1+k,q_2-k,q_3) + \right.
\nonumber \\
& & \left. \psi(q_1,q_2+k,q_3-k) + \psi(q_1-k,q_2,q_3+k) \right]  .   
\label{3UV1d:eq:three}
\end{eqnarray}
Here $\varepsilon(q) = - 2t \cos(q)$ is the free-particle spectrum, $N$ is the number of sites, and 
\begin{equation}
{\cal V}(k) =  U - 2 V \cos(k) \: , 
\label{3UV1d:eq:four}
\end{equation}
is the Fourier-transform of the pair-wise interaction potential. Because of the separable nature of ${\cal V}$, the right-hand-side of eq.~(\ref{3UV1d:eq:three}) contains a finite number of integrals of the form
\begin{equation}
F = \frac{1}{N} \sum_{k} f(k) \psi(k, q_2, q_1+q_3-k) \: , 
\label{3UV1d:eq:five}
\end{equation}
with various permutations of the arguments of $\psi$, various permutations of particle indices 1, 2, and 3, and with $f(k) = 1$, $\cos(k)$ or $\sin(k)$. Since the total momentum $P = q_1 + q_2 + q_3$ is a conserved quantity, $F$ can be rewritten as a function of one momentum only, in this example as a function of $q_2$
\begin{equation}
F = F(q_2) = \frac{1}{N} \sum_{k} f(k) \psi(k, q_2, P - q_2 - k) \: . 
\label{3UV1d:eq:six}
\end{equation}
Next, from the Schr\"odinger equation the wave function $\psi$ is expressed as a linear combination of $F$'s. Substituting $\psi$ back into definitions (\ref{3UV1d:eq:six}) results in a system of coupled integral equations for functions $F_{\alpha}(q)$. These equations have a general form
\begin{eqnarray}
& & \left[ \delta_{\alpha \beta} - 
\frac{1}{N} \sum_{k} \frac{A_{\alpha \beta}(k,q)}{E - \Delta(k,q)} \right] 
\left[ \begin{array}{c}
F_1(q) \\ F_2(q) \\ \vdots
\end{array} \right] =
\nonumber \\ 
& & \frac{1}{N} \sum_{k} \frac{B_{\alpha \beta}(k,q)}{E - \Delta(k,q)} 
\left[ \begin{array}{c}
F_1(k) \\ F_2(k) \\ \vdots
\end{array} \right] , 
\label{3UV1d:eq:seven}
\end{eqnarray}
where
\begin{equation}
\Delta(k,q) \equiv -2 t \, [ \cos(k) + \cos(q) + \cos(P-k-q) ] \: . 
\label{3UV1d:eq:eight}
\end{equation}
\begin{onecolumn}
Thus the original three-variable Schr\"odinger equation (\ref{3UV1d:eq:three}) is reduced to a system of one-variable integral equations~(\ref{3UV1d:eq:seven}). The latter can now be approximated by a finite-size matrix equation and trion energy $E$ deduced from the consistency condition. One should add that a similar reduction of the two-particle Schr\"odinger equation leads to a finite-dimensional system of algebraic equations, which has allowed analytical determination of the pair binding condition in several $UV$-like models~\cite{Micnas1990,Lin1991,Petukhov1992,Kornilovitch1993,Kornilovitch1995,daVeiga2002,Kornilovitch2004,Bak2007}.   

The number of functions $F$ in eq.~(\ref{3UV1d:eq:seven}) depends on the spatial dimensionality and radius of the interaction and on the wave function symmetry. For three distinguishable particles, there are 3 irreducible arrangements of $k$, $q$, and $(P-k-q)$ as arguments of $\psi$. For example, $\sum_{k} \psi(P-q-k,k,q) \neq \sum_{k} \psi(P-q-k,q,k)$. [This number is not 6 because under the integral $k$ can be converted to $(P-k-q)$ and vice versa by a change of variables.] In addition, the number of interactive sites in the one-dimensional $UV$ model is also 3. Thus in this general case, $3 \times 3 = 9$ functions $F$ are required. The number can be lowered if permutation symmetries of the wave function are taken into account. Not only it reduces computational complexity, but it also allows focusing numerical solution on particular trion states. 

{\em Total spin $S = 0$.} The wave function $\psi^0$ of spin-0 bosons is fully symmetrical with respect of permutation of all three arguments. As a result, functions $F$ can be defined as in eq.~(\ref{3UV1d:eq:six}), that is, with $\psi$ arguments arranged as $(k, q_i, P-q_i-k)$. Reduction from eq.~(\ref{3UV1d:eq:three}) to eq.~(\ref{3UV1d:eq:seven}) can be done with 3 functions
\begin{eqnarray}
F^{0}_1(q) & = & N^{-1} \sum_{k}         \, \psi^{0}(k,q,P-q-k) \: ,
\label{3UV1d:eq:nine} \\
F^{0}_2(q) & = & N^{-1} \sum_{k} \cos(k) \, \psi^{0}(k,q,P-q-k) \: ,
\label{3UV1d:eq:ten} \\
F^{0}_3(q) & = & N^{-1} \sum_{k} \sin(k) \, \psi^{0}(k,q,P-q-k) \: .
\label{3UV1d:eq:eleven}
\end{eqnarray}
The matrices $A^{0}_{\alpha \beta}$ and $B^{0}_{\alpha \beta}$ for the $S = 0$ trion are 
\begin{equation}
A^{0}_{\alpha \beta} = \left[ \begin{array}{ccc}
U           & -V           ({\rm c}_k + {\rm c}_P) & -V           ({\rm s}_k + {\rm s}_P)  \\
U {\rm c}_k & -V {\rm c}_k ({\rm c}_k + {\rm c}_P) & -V {\rm c}_k ({\rm s}_k + {\rm s}_P)  \\
U {\rm s}_k & -V {\rm s}_k ({\rm c}_k + {\rm c}_P) & -V {\rm s}_k ({\rm s}_k + {\rm s}_P)  
\end{array} \right] ,
\label{3UV1d:eq:twelve}
\end{equation}
\begin{equation}
B^{0}_{\alpha \beta} = \left[ 
\begin{array}{ccc}
 2U                                               & 
-2V                    ({\rm c}_q \! + {\rm c}_{\! P})      & 
-2V                    ({\rm s}_q \! + {\rm s}_{\! P})  \\
 U ({\rm c}_k \! +{\rm c}_{\! P})                            & 
\hspace{-0.2cm} -V ({\rm c}_k \! + {\rm c}_{\! P}) ({\rm c}_q \! + {\rm c}_{\! P}) & 
\hspace{-0.2cm} -V ({\rm c}_k \! + {\rm c}_{\! P}) ({\rm s}_q \! + {\rm s}_{\! P})  \\
 U ({\rm s}_k \! +{\rm s}_{\! P})                            & 
\hspace{-0.2cm} -V ({\rm s}_k \! + {\rm s}_{\! P}) ({\rm c}_q \! + {\rm c}_{\! P})  & 
\hspace{-0.2cm} -V ({\rm s}_k \! + {\rm s}_{\! P}) ({\rm s}_q \! + {\rm s}_{\! P})  
\end{array}  
\right] ,
\label{3UV1d:eq:thirteen}
\end{equation}
where the following shorthand notation has been adopted: ${\rm c}_{k,q} \equiv \cos(k,q)$, ${\rm s}_{k,q} \equiv \sin(k,q)$, ${\rm c}_{P} \equiv \cos(P-q-k)$, ${\rm s}_{P} \equiv \sin(P-q-k)$. In this form, eigenvalue equation~(\ref{3UV1d:eq:seven}) is equivalent to the one given by Rudin in ref.~\cite{Rudin1986}. 

One should note that in fact functions $F^0_{2}$ and $F^0_{3}$ are not independent. Changing variable $k' = P-q-k$ and making use of permutation symmetry, one can show that 
\begin{equation}
[ 1 - \cos{(P-q)} ] F^{0}_{2}(q) = \sin{(P-q)} F^{0}_{3}(q) \: .
\label{3UV1d:eq:Flink}
\end{equation}
As a result, the size of matrices $A$ and $B$ can be reduced further from 3 to 2. Unfortunately, relation (\ref{3UV1d:eq:Flink}) generates singular kernels in eq.~(\ref{3UV1d:eq:seven}), and for this reason is not employed in this work. Instead, relations like (\ref{3UV1d:eq:Flink}) are used to ensure the self-consistency of the numerical solution of the eigenvalue equation (\ref{3UV1d:eq:seven}).  

{\em Total spin $S = 1/2$.} There are two equivalent ways to access the two $S = 1/2$ states: by symmetrizing or antisymmetrizing the wave functions with respect to permutation of two coordinates, for example $q_1$ and $q_2$. (In the $S = 1/2$ sector, it is not possible to choose basis states to be eigenfunctions of all three permutation operators.) In the former case, eigenvalue equation (\ref{3UV1d:eq:seven}) will also contain the fully symmetrical solution ($S = 0$) as a subset, while in the latter case it will contain the fully antisymmetrical solution ($S = 3/2$) as a subset. We choose to work with antisymmetrical states $\psi_{-}(q_1,q_2,q_3) = - \psi_{-}(q_2,q_1,q_3)$ because with this choice an $S = 1/2$ trion is the lowest energy solution. Two irreducible arrangements of $\psi_{-}$ arguments, times three interacting sites, makes 6 possible combinations, of which one vanishes due to the $(q_1 \leftrightarrow q_2)$ antisymmetry. Reduction to eq.~(\ref{3UV1d:eq:seven}) requires 5 functions:  
\begin{eqnarray}
F^{1/2}_1(q) & = & N^{-1} \sum_{k}         \, \psi_{-}(k,q,P-q-k) \, ,
\label{3UV1d:eq:fourteen}  \\
F^{1/2}_2(q) & = & N^{-1} \sum_{k} \cos(k) \, \psi_{-}(k,P-q-k,q) \, ,
\label{3UV1d:eq:fifteen}   \\
F^{1/2}_3(q) & = & N^{-1} \sum_{k} \cos(k) \, \psi_{-}(k,q,P-q-k) \, ,
\label{3UV1d:eq:sixteen}   \\
F^{1/2}_4(q) & = & N^{-1} \sum_{k} \sin(k) \, \psi_{-}(k,P-q-k,q) \, ,
\label{3UV1d:eq:seventeen} \\
F^{1/2}_5(q) & = & N^{-1} \sum_{k} \sin(k) \, \psi_{-}(k,q,P-q-k) \, .
\label{3UV1d:eq:eighteen}
\end{eqnarray}
The resulting system (\ref{3UV1d:eq:seven}) is defined by $5 \times 5$ matrices $A$ and $B$:
\begin{equation}
A^{1/2}_{\alpha \beta} = \left[  
\begin{array}{ccccc} 
 U            &  0     & -2V {\rm c}_{k}   & 0 & -2V {\rm s}_{k}             \\
 0            & -V {\rm c}_{k} ({\rm c}_{k} - {\rm c}_{P}) &  0  
              & -V {\rm c}_{k} ({\rm s}_{k} - {\rm s}_{P}) &  0              \\
U {\rm c}_{k} &  0     & -2V {\rm c}^2_{k} & 0 & -2V {\rm c}_{k} {\rm s}_{k} \\
 0            & -V {\rm s}_{k} ({\rm c}_{k} - {\rm c}_{P}) &  0 
              & -V {\rm s}_{k} ({\rm s}_{k} - {\rm s}_{P}) &  0              \\
U {\rm s}_{k} &  0     & -2V {\rm s}_{k} {\rm c}_{k} & 0 & -2V {\rm s}^2_{k} 
\end{array} \right] \: ,
\label{3UV1d:eq:nineteen}
\end{equation}
\begin{equation}
B^{1/2}_{\alpha \beta} = \left[  
\begin{array}{ccccc} 
-U             & -V ({\rm c}_{P} - {\rm c}_{q}) & -2V (- {\rm c}_{q})  
               & -V ({\rm s}_{P} - {\rm s}_{q}) & -2V (- {\rm s}_{q}) \\
 U ({\rm c}_{P} - {\rm c}_{k}) 
               & 0 & -2V {\rm c}_{P}({\rm c}_{P} - {\rm c}_{k}) 
               & 0 & -2V {\rm s}_{P}({\rm c}_{P} - {\rm c}_{k}) \\
-U {\rm c}_{k} & -V {\rm c}_{P}({\rm c}_{P} - {\rm c}_{q}) 
               & -2V (-{\rm c}_{k}{\rm c}_{q}) 
               & -V {\rm c}_{P}({\rm s}_{P} - {\rm s}_{q})  
               & -2V (-{\rm c}_{k}{\rm s}_{q}) \\
 U ({\rm s}_{P} - {\rm s}_{k}) 
               & 0 & -2V {\rm c}_{P}({\rm s}_{P} - {\rm s}_{k})  
               & 0 & -2V {\rm s}_{P}({\rm s}_{P} - {\rm s}_{k}) \\
-U {\rm s}_{k} & -V {\rm s}_{P}({\rm c}_{P} - {\rm c}_{q})  
               & -2V (-{\rm s}_{k}{\rm c}_{q}) 
               & -V {\rm s}_{P}({\rm s}_{P} - {\rm s}_{q})   
               & -2V (-{\rm s}_{k}{\rm s}_{q}) 
\end{array} \right] \: .
\label{3UV1d:eq:twenty}
\end{equation}

{\em Total spin $S = 3/2$.} This sector of spin-$\frac{1}{2}$ fermions is characterized by the three-row Young diagram and fully antisymmetrical coordinate wave functions. One irreducible arrangement of $\psi$ arguments, times three interacting sites, makes 3 combinations, of which one vanishes. Only 2 functions are needed to complete the reduction: 
\begin{eqnarray}
F^{3/2}_1(q) & = & N^{-1} \sum_{k} \cos(k) \, \psi^{3/2}(k,q,P-q-k) \, ,
\label{3UV1d:eq:twentyone}  \\
F^{3/2}_2(q) & = & N^{-1} \sum_{k} \sin(k) \, \psi^{3/2}(k,q,P-q-k) \, .
\label{3UV1d:eq:twentytwo}   
\end{eqnarray}
The eigenvalue equation (\ref{3UV1d:eq:seven}) is defined by $2 \times 2$ matrices
\begin{equation}
A^{3/2}_{\alpha \beta} = \left[  
\begin{array}{cc}
- V {\rm c}_k ( {\rm c}_k - {\rm c}_P ) & -V {\rm c}_k ( {\rm s}_k - {\rm s}_P ) \\
- V {\rm s}_k ( {\rm c}_k - {\rm c}_P ) & -V {\rm s}_k ( {\rm s}_k - {\rm s}_P ) 
\end{array} \right] ,
\label{3UV1d:eq:twentythree}
\end{equation}
\begin{equation}
B^{3/2}_{\alpha \beta} = \left[   
\begin{array}{cc}
- V ( {\rm c}_k - {\rm c}_P )( {\rm c}_P - {\rm c}_q ) & 
\!\! - V ( {\rm c}_k - {\rm c}_P )( {\rm s}_P - {\rm s}_q ) \\
- V ( {\rm s}_k - {\rm s}_P )( {\rm c}_P - {\rm c}_q ) & 
\!\! - V ( {\rm s}_k - {\rm s}_P )( {\rm s}_P - {\rm s}_q )
\end{array} \right] .
\label{3UV1d:eq:twentyfour}
\end{equation}
As mentioned above, trion energies of the $S = 3/2$ sector must coincide with one of the excited states of the $S = 1/2$ sector. Such a coincidence serves as a consistency check of the entire method. 

To obtain trion energies, $k$-integrals in eq.~(\ref{3UV1d:eq:seven}) are approximated by discrete sums using the Simpson rule. The absence of Efimov effect in 1D ensures that this procedure is well-defined. Care must be taken in treating the singularity at $(k,q) = 0$ at small binding energies. This is addressed by using a denser mesh near $k = 0$. The number of $k$ points is increased until convergence is achieved. The bulk of the results presented below have been obtained with 60 $k$-points in the $[-\pi,\pi]$ interval. Once eq.~(\ref{3UV1d:eq:seven}) is converted to a finite-size matrix equation, eigenvalues $\lambda_i$ are computed and energy $E$ is adjusted to determine when an eigenvalue crosses $\lambda = 1$. Each crossing corresponds to a trion state. The procedure is repeated for every $(U, V, P)$ combination.

\section{\label{3UV1d:sec:four}
Results and discussion}

A typical spin-$\frac{1}{2}$ fermion trion spectrum is shown in fig.~\ref{3UV1d:fig:one}. The model parameters, $U = 20 \, t$ and $V = 2.1 \, t$, are close to the binding threshold. As a result, the binding energies are small, of the order of several tenths of $t$. The $S = 3/2$ trion has a higher energy (dashed line) than the two $S = 1/2$ trions (thin solid lines). Also shown are the bottom of the three-free-particle continuum $\{ 111 \}$ (dotted line) and the bottom of the singlet-pair plus a free particle continuum $\{ 21 \}$ (thick solid line). Notice that at $P = 0$, only the lowest $S = 1/2$ trion lies below the $\{21\}$ continuum. The other two states are unstable relative to decay to a singlet pair plus a free fermion. Near the Brillouin zone edges, all trion states are stable relative to the decay to constituent particles. A similar effect was observed in two-body problems~\cite{daVeiga2002,Kornilovitch2004}. Enhanced stability of bound complexes at large momenta might be a general feature of lattice models with attractive potentials.      

Figure~\ref{3UV1d:fig:two} shows variation of spin-$\frac{1}{2}$ fermion energies with $V$ for fixed on-site repulsion $U = 20 \, t$ and total momentum $P = 0$. The singlet pair forms at $V = 2Ut/(U+4t) = \frac{5}{3} \, t$; at this point the $\{21\}$ energy (thick solid line) separates from the $\{111\}$ continuum (dotted line). As $V$ increases further, trion states separate from the $\{21\}$ continuum: the first $S = 1/2$ trion at $V = 1.740 \, t$, the second $S = 1/2$ trion at $V = 2.129 \, t$, and the $S = 3/2$ trion at $V = 2.331 \,t$.

\end{onecolumn}
\begin{twocolumn}
\begin{figure}[t]
\begin{center}
\includegraphics[width=0.48\textwidth]{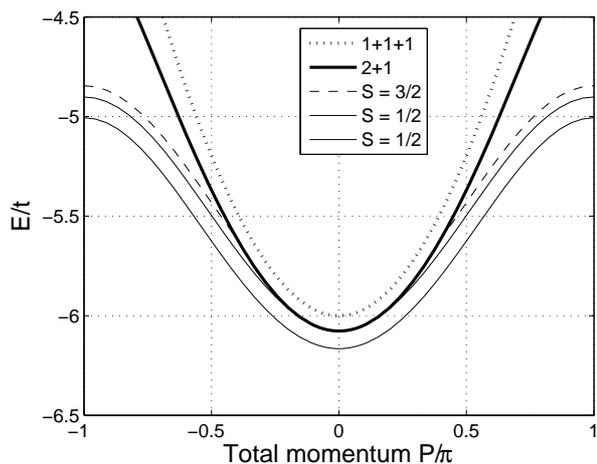}
\end{center}
\vspace{-0.4cm}
\caption{Energies of three spin-$\frac{1}{2}$ fermions for $U = 20 \, t$ and $V = 2.1 \, t$. The dotted line is the bottom of $\{ 111 \}$ free particle continuum, $E_{111} = -6t \cos{(P/3)}$, the thick solid line is the bottom of the $\{ 21 \}$ continuum. Note that the top $S = 1/2$ trion and the $S = 3/2$ trion are stable near the edges of the Brillouin zone but become unstable at small $P$.} 
\label{3UV1d:fig:one}
\end{figure}
\begin{figure}[t]
\begin{center}
\includegraphics[width=0.48\textwidth]{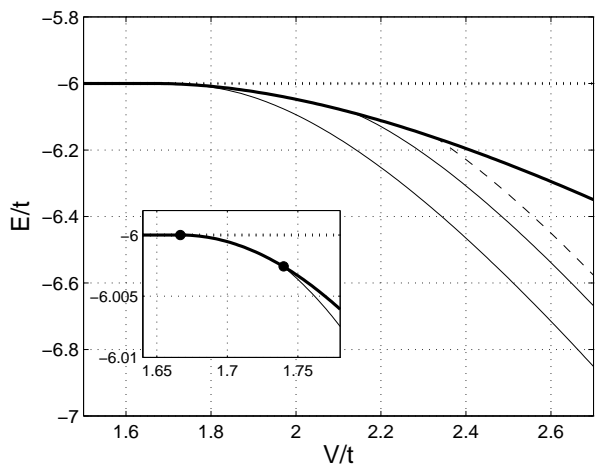}
\end{center}
\vspace{-0.4cm}
\caption{Fermion energy {\em vs.} attraction parameter $V$, for $U = 20 \, t$ and $P = 0$. The legend is the same as in fig.~\ref{3UV1d:fig:one}. The $\{21\}$ continuum (thick solid line) separates from $\{111\}$ free particle continuum at $V = \frac{5}{3}\, t$. Two $S = 1/2$ trions separate from $\{21\}$ at $V = 1.740 \, t$ and $V = 2.129 \, t$, respectively. $S = 3/2$ trion separates from $\{21\}$ at $V = 2.331 \, t$. Inset: an enlarged view of pair and $S = 1/2$ trion separation region. Separation points are marked with circles.} 
\label{3UV1d:fig:two}
\end{figure}

By repeating the procedure at all $U$, an entire phase diagram of the three-fermion $UV$ model can be constructed. It is shown in fig.~\ref{3UV1d:fig:three} for the most interesting case of $P = 0$. (In deriving the phase boundaries, a trion energy is compared to $E_2 - 2t$, where $E_2$ is the energy of a singlet fermion pair at zero pair momentum, and $-2t$ represents the energy of a free fermion.) The phase diagram contains a narrow but finite parameter region that separates formation of the singlet fermion pair and formation of the first $S = 1/2$ trion. Between the thick and the first thin solid lines, pairs form but trions do not. The ground state of the three--spin-$\frac{1}{2}$--fermion system is one singlet pair plus one free particle. The width of this region is shown in the inset; at $U > 10 \, t$ it scales as $\Delta V \approx 0.31 \, t/\sqrt{U/t}$. For $U \rightarrow \infty$, all phase boundaries converge to a critical attraction value $V = 2t$.

\begin{figure}[t]
\begin{center}
\includegraphics[width=0.48\textwidth]{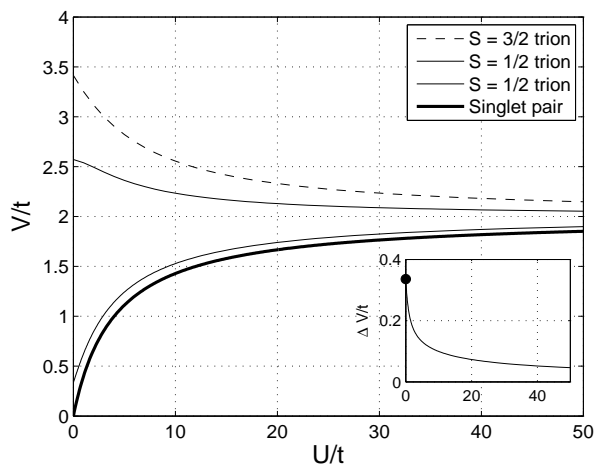}
\end{center}
\vspace{-0.4cm}
\caption{Main panel: phase diagram of three fermions on the one-dimensional $UV$ chain for $P = 0$. Bound complexes form above the respective lines. The thick solid line (singlet pair) is $V = 2Ut/(U + 4t)$. The two thin solid lines are $S = 1/2$ trions, the lower line starts at $U = 0$, $V = 0.336 \, t$. Inset: difference between the lower thin solid and thick solid lines: the width of $\{ 21 \}$ stability region.}
\label{3UV1d:fig:three}
\end{figure}

The existence of a parameter region where fermion pairs form while trions do not has implications for superconductivity. The preformed real-space pairs scenario~\cite{Micnas1990,Alexandrov1994,Alexandrov2010} posits an attractive interaction strong enough to overcome direct Coulomb repulsion between the particles, at least at a finite separation. The attraction could be of phonon, nonphonon or mixed origin. Clustering is an obvious concern for those theories. If the attraction is so strong, why would it not pull all the particles into one macroscopic cluster, which would then prevent any particle mobility and hence superconductivity? In the case of zero-range attraction the exclusion principle prevents particles from feeling more than one attractive bond at a time, and clustering does not occur~\cite{Takahashi1970,Mattis1984,Rudin1985}. The results of the present paper suggest that regions of pair stability exist even for finite-range potentials. Adding a third fermion to an existing bound pair requires anti-symmetrizing the wave function with respect to at least two coordinates, which implies an infinite on-site repulsion in that channel. If the attraction is barely above the pair-forming threshold (for a finite $U$) it may not be strong enough to overcome an effective infinite $U$ brought by the third fermion, so the trion does not form. This reasoning also helps understand why the pair stability region shrinks at larger $U$: there is less and less difference between the real dynamical $U$ and the effective infinite $U$ of the exclusion principle. This argument is general and should also hold for other shapes of (repulsive core)-(attractive shell) potentials in one and two dimensions, although details will differ. 

\begin{figure}[t]
\begin{center}
\includegraphics[width=0.48\textwidth]{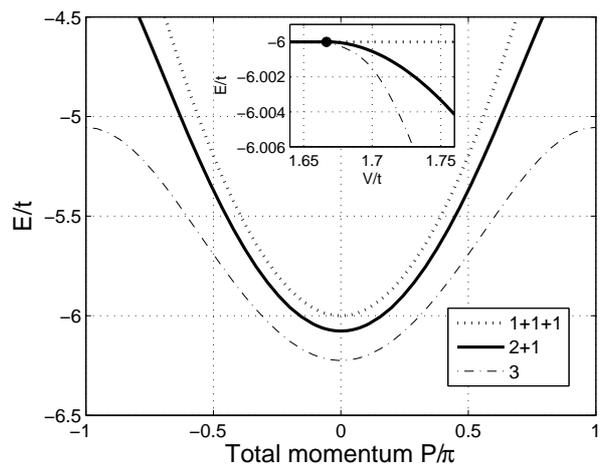}
\end{center}
\vspace{-0.4cm}
\caption{Energies of three spin-0 bosons. Main panel: spectrum of $S = 0$ trion is compared with those of three free bosons and boson-pair-plus-free-boson systems, for $U = 20 \, t$ and $V = 2.1 \, t$. Inset: energy {\em vs.} $V$ for $U = 20 \, t$ and $P = 0$.}
\label{3UV1d:fig:five}
\end{figure}

Energies of three spin-0 bosons are presented in fig.~\ref{3UV1d:fig:five}. The $S = 0$ trion always has a lower energy than the fermion $S = 1/2$ trion for the same $U$, $V$, and $P$. The $S = 0$ trion separates from the $\{ 111 \}$ free particle continuum at the same $V$ as the boson pair, see inset. This suggests that the boson pairs and trions may have the same binding thresholds.

\section{\label{3UV1d:sec:five}
Summary}

The three-body quantum-mechanical problem has been studied on a one-dimensional lattice with on-site repulsion and nearest-neighbour attraction. Making use of momentum conservation and separability of the potential, the Schr\"odinger equation has been reduced to a system of coupled one-variable integral equations (\ref{3UV1d:eq:seven}). The latter has been approximated by a matrix equation and solved numerically. Eigenvalue equations (\ref{3UV1d:eq:seven}) have been derived for fully symmetrical wave functions (spin-0 bosons, total spin $S=0$), fully antisymmetrical wave functions (polarized spin-$\frac{1}{2}$ fermions, total spin $S = 3/2$), and mixed-symmetry wave functions (ground state of spin-$\frac{1}{2}$ fermions, total spin $S = 1/2$). Trion energies have been computed in all three spin sectors. Phase boundaries have been derived by comparing energies of bound trions with energies of bound pairs. A narrow but finite parameter region has been identified where the ground state of spin-$\frac{1}{2}$ fermions is one bound singlet pair plus one free particle.

\acknowledgments

The author wishes to thank Sasha Alexandrov for providing inspiration for this work, and Vladimir Bulatov, James Hague, and Gerardo Sica for useful discussions on the subject of this paper.

\end{twocolumn}

\newpage

\begin{onecolumn}

SUPPLEMENTAL MATERIAL: EXPANDED VERSIONS OF INTEGRAL EQUATIONS (\ref{3UV1d:eq:seven}).

\vspace{0.5cm}

{\em Total spin} $S = 0$. Three functions $F^{0}_1(q)$, $F^{0}_2(q)$, and $F^{0}_3(q)$ are defined in eqs.~(\ref{3UV1d:eq:nine})-(\ref{3UV1d:eq:eleven}). The eigenvalue equation (\ref{3UV1d:eq:seven}) is
\begin{eqnarray}
F^{0}_1(q) & = & \frac{U}{N} \sum_{k} 
\frac{F^{0}_1(q) + 2 F^{0}_1(k)}
{E - \varepsilon(k) - \varepsilon(q) - \varepsilon(P-q-k)}
\nonumber \\
       &   & - \frac{V}{N} \sum_{k}
\frac{[\cos(k) + \cos(P-q-k)] F^{0}_2(q)}
{E - \varepsilon(k) - \varepsilon(q) - \varepsilon(P-q-k)}      
\nonumber \\
       &   & - \frac{V}{N} \sum_{k}
\frac{2 [\cos(q) + \cos(P-q-k)] F^{0}_2(k)}
{E - \varepsilon(k) - \varepsilon(q) - \varepsilon(P-q-k)}      
\nonumber \\
       &   & - \frac{V}{N} \sum_{k}
\frac{[\sin(k) + \sin(P-q-k)] F^{0}_3(q)}
{E - \varepsilon(k) - \varepsilon(q) - \varepsilon(P-q-k)}
\nonumber \\
       &   & - \frac{V}{N} \sum_{k}
\frac{2 [\sin(q) + \sin(P-q-k)] F^{0}_3(k)}
{E - \varepsilon(k) - \varepsilon(q) - \varepsilon(P-q-k)}     \: , 
\nonumber 
\end{eqnarray}
\begin{eqnarray}
F^{0}_2(q) & = & \frac{U}{N} \sum_{k} 
\frac{\cos(k) F^{0}_1(q) + [\cos(k) + \cos(P-q-k)] F^{0}_1(k)}
{E - \varepsilon(k) - \varepsilon(q) - \varepsilon(P-q-k)}
\nonumber \\
       &   & - \frac{V}{N} \sum_{k}
\frac{\cos(k)[\cos(k)+\cos(P-q-k)] F^{0}_2(q)}
{E - \varepsilon(k) - \varepsilon(q) - \varepsilon(P-q-k)}      
\nonumber \\
       &   & - \frac{V}{N} \sum_{k}
\frac{[\cos(k)+\cos(P-q-k)][\cos(q)+\cos(P-q-k)] F^{0}_2(k)}
{E - \varepsilon(k) - \varepsilon(q) - \varepsilon(P-q-k)}      
\nonumber \\
       &   & - \frac{V}{N} \sum_{k}
\frac{\cos(k)[\sin(k)+\sin(P-q-k)] F^{0}_3(q)}
{E - \varepsilon(k) - \varepsilon(q) - \varepsilon(P-q-k)}
\nonumber \\
       &   & - \frac{V}{N} \sum_{k}
\frac{[\cos(k)+\cos(P-q-k)][\sin(q)+\sin(P-q-k)] F^{0}_3(k)}
{E - \varepsilon(k) - \varepsilon(q) - \varepsilon(P-q-k)}     \: , 
\nonumber 
\end{eqnarray}
\begin{eqnarray}
F^{0}_3(q) & = & \frac{U}{N} \sum_{k} 
\frac{\sin(k) F^{0}_1(q) + [\sin(k) + \sin(P-q-k)] F^{0}_1(k)}
{E - \varepsilon(k) - \varepsilon(q) - \varepsilon(P-q-k)}
\nonumber \\
       &   & - \frac{V}{N} \sum_{k}
\frac{\sin(k)[\cos(k)+\cos(P-q-k)] F^{0}_2(q)}
{E - \varepsilon(k) - \varepsilon(q) - \varepsilon(P-q-k)}      
\nonumber \\
       &   & - \frac{V}{N} \sum_{k}
\frac{[\sin(k)+\sin(P-q-k)][\cos(q)+\cos(P-q-k)] F^{0}_2(k)}
{E - \varepsilon(k) - \varepsilon(q) - \varepsilon(P-q-k)}      
\nonumber \\
       &   & - \frac{V}{N} \sum_{k}
\frac{\sin(k)[\sin(k)+\sin(P-q-k)] F^{0}_3(q)}
{E - \varepsilon(k) - \varepsilon(q) - \varepsilon(P-q-k)}
\nonumber \\
       &   & - \frac{V}{N} \sum_{k}
\frac{[\sin(k)+\sin(P-q-k)][\sin(q)+\sin(P-q-k)] F^{0}_3(k)}
{E - \varepsilon(k) - \varepsilon(q) - \varepsilon(P-q-k)}     \: . 
\nonumber
\end{eqnarray}

\vspace{1.0cm}

{\em Total spin} $S = 1/2$. Five functions $F^{1/2}_1(q)$, $F^{1/2}_2(q)$, $F^{1/2}_3(q)$, $F^{1/2}_4(q)$ and $F^{1/2}_5(q)$ are defined in eqs.~(\ref{3UV1d:eq:fourteen})-(\ref{3UV1d:eq:eighteen}). The eigenvalue equation (\ref{3UV1d:eq:seven}) is 
\begin{eqnarray}
F^{1/2}_1(q) & = & \frac{U}{N} \sum_{k} 
\frac{F^{1/2}_1(q) - F^{1/2}_1(k)}
{E - \varepsilon(k) - \varepsilon(q) - \varepsilon(P-q-k)}
\nonumber \\
       &   & - \frac{V}{N} \sum_{k}
\frac{[\cos(P-q-k) - \cos(q)] F^{1/2}_2(k)}
{E - \varepsilon(k) - \varepsilon(q) - \varepsilon(P-q-k)}      
\nonumber \\
       &   & - \frac{V}{N} \sum_{k}
\frac{ 2\cos(k) F^{1/2}_3(q) - 2\cos(q) F^{0}_3(k)}
{E - \varepsilon(k) - \varepsilon(q) - \varepsilon(P-q-k)}      
\nonumber \\
       &   & - \frac{V}{N} \sum_{k}
\frac{[\sin(P-q-k) - \sin(q)] F^{1/2}_4(k)}
{E - \varepsilon(k) - \varepsilon(q) - \varepsilon(P-q-k)}      
\nonumber \\
       &   & - \frac{V}{N} \sum_{k}
\frac{ 2\sin(k) F^{1/2}_5(q) - 2\sin(q) F^{0}_5(k)}
{E - \varepsilon(k) - \varepsilon(q) - \varepsilon(P-q-k)} \: ,     
\nonumber
\end{eqnarray}
\begin{eqnarray}
F^{1/2}_2(q) & = & \frac{U}{N} \sum_{k} 
\frac{[\cos(P-q-k) - \cos(k)] F^{1/2}_1(k)}
{E - \varepsilon(k) - \varepsilon(q) - \varepsilon(P-q-k)}
\nonumber \\
       &   & - \frac{V}{N} \sum_{k}
\frac{\cos(k) [\cos(k) - \cos(P-q-k)] F^{1/2}_2(q)}
{E - \varepsilon(k) - \varepsilon(q) - \varepsilon(P-q-k)}      
\nonumber \\
       &   & - \frac{V}{N} \sum_{k}
\frac{ 2\cos(P-q-k) [ \cos(P-q-k) - \cos(k)] F^{0}_3(k)}
{E - \varepsilon(k) - \varepsilon(q) - \varepsilon(P-q-k)}      
\nonumber \\
       &   & - \frac{V}{N} \sum_{k}
\frac{\cos(k)[\sin(k) - \sin(P-q-k)] F^{1/2}_4(q)}
{E - \varepsilon(k) - \varepsilon(q) - \varepsilon(P-q-k)}      
\nonumber \\
       &   & - \frac{V}{N} \sum_{k}
\frac{ 2\sin(P-q-k) [ \cos(P-q-k) - \cos(k)] F^{0}_5(k)}
{E - \varepsilon(k) - \varepsilon(q) - \varepsilon(P-q-k)} \: ,     
\nonumber
\end{eqnarray}
\begin{eqnarray}
F^{1/2}_3(q) & = & \frac{U}{N} \sum_{k} 
\frac{\cos(k) F^{1/2}_1(q) - \cos(k) F^{1/2}_1(k)}
{E - \varepsilon(k) - \varepsilon(q) - \varepsilon(P-q-k)}
\nonumber \\
       &   & - \frac{V}{N} \sum_{k}
\frac{\cos(P-q-k) [ \cos(P-q-k) - \cos(q) ] F^{1/2}_2(k)}
{E - \varepsilon(k) - \varepsilon(q) - \varepsilon(P-q-k)}      
\nonumber \\
       &   & - \frac{V}{N} \sum_{k}
\frac{ 2\cos^2(k) F^{0}_3(q) - 2\cos(k)\cos(q) F^{0}_3(k) }
{E - \varepsilon(k) - \varepsilon(q) - \varepsilon(P-q-k)}      
\nonumber \\
       &   & - \frac{V}{N} \sum_{k}
\frac{\cos(P-q-k)[ \sin(P-q-k) - \sin(q) ] F^{1/2}_4(k)}
{E - \varepsilon(k) - \varepsilon(q) - \varepsilon(P-q-k)}      
\nonumber \\
       &   & - \frac{V}{N} \sum_{k}
\frac{ 2 \cos(k)\sin(k) F^{0}_5(q) - 2 \cos(k)\sin(q) F^{0}_5(k) }
{E - \varepsilon(k) - \varepsilon(q) - \varepsilon(P-q-k)} \: ,     
\nonumber
\end{eqnarray}
\begin{eqnarray}
F^{1/2}_4(q) & = & \frac{U}{N} \sum_{k} 
\frac{[\sin(P-q-k) - \sin(k)] F^{1/2}_1(k)}
{E - \varepsilon(k) - \varepsilon(q) - \varepsilon(P-q-k)}
\nonumber \\
       &   & - \frac{V}{N} \sum_{k}
\frac{\sin(k) [\cos(k) - \cos(P-q-k)] F^{1/2}_2(q)}
{E - \varepsilon(k) - \varepsilon(q) - \varepsilon(P-q-k)}      
\nonumber \\
       &   & - \frac{V}{N} \sum_{k}
\frac{ 2\cos(P-q-k) [ \sin(P-q-k) - \sin(k)] F^{0}_3(k)}
{E - \varepsilon(k) - \varepsilon(q) - \varepsilon(P-q-k)}      
\nonumber \\
       &   & - \frac{V}{N} \sum_{k}
\frac{\sin(k)[\sin(k) - \sin(P-q-k)] F^{1/2}_4(q)}
{E - \varepsilon(k) - \varepsilon(q) - \varepsilon(P-q-k)}      
\nonumber \\
       &   & - \frac{V}{N} \sum_{k}
\frac{ 2\sin(P-q-k) [ \sin(P-q-k) - \sin(k)] F^{0}_5(k)}
{E - \varepsilon(k) - \varepsilon(q) - \varepsilon(P-q-k)} \: ,     
\nonumber
\end{eqnarray}
\begin{eqnarray}
F^{1/2}_5(q) & = & \frac{U}{N} \sum_{k} 
\frac{\sin(k) F^{1/2}_1(q) - \sin(k) F^{1/2}_1(k)}
{E - \varepsilon(k) - \varepsilon(q) - \varepsilon(P-q-k)}
\nonumber \\
       &   & - \frac{V}{N} \sum_{k}
\frac{\sin(P-q-k) [ \cos(P-q-k) - \cos(q) ] F^{1/2}_2(k)}
{E - \varepsilon(k) - \varepsilon(q) - \varepsilon(P-q-k)}      
\nonumber \\
       &   & - \frac{V}{N} \sum_{k}
\frac{ 2\sin(k)\cos(k) F^{0}_3(q) - 2\sin(k)\cos(q) F^{0}_3(k) }
{E - \varepsilon(k) - \varepsilon(q) - \varepsilon(P-q-k)}      
\nonumber \\
       &   & - \frac{V}{N} \sum_{k}
\frac{\sin(P-q-k)[ \sin(P-q-k) - \sin(q) ] F^{1/2}_4(k)}
{E - \varepsilon(k) - \varepsilon(q) - \varepsilon(P-q-k)}      
\nonumber \\
       &   & - \frac{V}{N} \sum_{k}
\frac{ 2 \sin^2(k) F^{0}_5(q) - 2 \sin(k)\sin(q) F^{0}_5(k) }
{E - \varepsilon(k) - \varepsilon(q) - \varepsilon(P-q-k)} \: .     
\nonumber
\end{eqnarray}

\vspace{1.0cm}

{\em Total spin} $S = 3/2$. Two functions $F^{3/2}_1(q)$ and $F^{3/2}_2(q)$ are defined in eqs.~(\ref{3UV1d:eq:twentyone}) and (\ref{3UV1d:eq:twentytwo}). The eigenvalue equation (\ref{3UV1d:eq:seven}) is 
\begin{eqnarray}
F^{3/2}_1(q) & = & - \frac{V}{N} \sum_{k}
\frac{\cos(k) [ \cos(k) - \cos(P-q-k) ] F^{3/2}_1(q)}
{E - \varepsilon(k) - \varepsilon(q) - \varepsilon(P-q-k)}      
\nonumber \\
       &   & - \frac{V}{N} \sum_{k}
\frac{ [\cos(k) - \cos(P-q-k)] [\cos(P-q-k) - \cos(q)] F^{3/2}_1(k) }
{E - \varepsilon(k) - \varepsilon(q) - \varepsilon(P-q-k)}      
\nonumber \\
       &   & - \frac{V}{N} \sum_{k}
\frac{\cos(k) [ \sin(k) - \sin(P-q-k) ] F^{3/2}_2(q)}
{E - \varepsilon(k) - \varepsilon(q) - \varepsilon(P-q-k)}      
\nonumber \\
       &   & - \frac{V}{N} \sum_{k}
\frac{ [\cos(k) - \cos(P-q-k)] [\sin(P-q-k) - \sin(q)] F^{3/2}_2(k) }
{E - \varepsilon(k) - \varepsilon(q) - \varepsilon(P-q-k)}  \: ,           
\nonumber
\end{eqnarray}
\begin{eqnarray}
F^{3/2}_2(q) & = & - \frac{V}{N} \sum_{k}
\frac{\sin(k) [ \cos(k) - \cos(P-q-k) ] F^{3/2}_1(q)}
{E - \varepsilon(k) - \varepsilon(q) - \varepsilon(P-q-k)}      
\nonumber \\
       &   & - \frac{V}{N} \sum_{k}
\frac{ [\sin(k) - \sin(P-q-k)] [\cos(P-q-k) - \cos(q)] F^{3/2}_1(k) }
{E - \varepsilon(k) - \varepsilon(q) - \varepsilon(P-q-k)}      
\nonumber \\
       &   & - \frac{V}{N} \sum_{k}
\frac{\sin(k) [ \sin(k) - \sin(P-q-k) ] F^{3/2}_2(q)}
{E - \varepsilon(k) - \varepsilon(q) - \varepsilon(P-q-k)}      
\nonumber \\
       &   & - \frac{V}{N} \sum_{k}
\frac{ [\sin(k) - \sin(P-q-k)] [\sin(P-q-k) - \sin(q)] F^{3/2}_2(k) }
{E - \varepsilon(k) - \varepsilon(q) - \varepsilon(P-q-k)}  \: .           
\nonumber
\end{eqnarray}

\end{onecolumn}

\end{document}